\documentclass[
 reprint,
 amsmath,
 amssymb,
 aps,
 superscriptaddress,
 longbibliography,
]{revtex4-2}

\usepackage{CJKutf8}
\usepackage[usenames,dvipsnames,table]{xcolor}
\usepackage[pdftex]{graphicx}

\usepackage{hyperref}
\usepackage{booktabs}
\usepackage[normalem]{ulem}
\hypersetup{
    colorlinks=true,       
    linkcolor=blue,          
    citecolor=magenta,        
    filecolor=red,      
    urlcolor=cyan           
}

\usepackage{lmodern}

\usepackage{physics}
\usepackage{amsmath}
\usepackage{xcolor}

\begin{document}

\title{Direct and inverse cascades scaling in real shell models of turbulence }

\author{James Creswell}
\affiliation{
Ludwig Maximilian University,
Theresienstr.\ 37, 80333 Munich, Germany
}

\author{Viatcheslav Mukhanov}%
\affiliation{
Ludwig Maximilian University of Munich,
Theresienstr.\ 37, 80333 Munich, Germany
}

\author{Yaron Oz}%
\affiliation{
School of Physics and Astronomy, Tel-Aviv University, Tel-Aviv 69978, Israel
}
%

%
\date{\today}

\begin{abstract}
Shell models provide a simplified mathematical framework that captures essential features of incompressible fluid turbulence,
such as the energy cascade and scaling of the fluid observables.
We perform a precision analysis of the direct and inverse cascades  
in shell models of turbulence, where the velocity field is a real-valued function. 
We calculate the leading hundred anomalous scaling exponents, 
the marginal probability distribution functions of the velocity field at different shells, as well as the correlations between
different shells. We find that the structure functions in both cascades exhibit a linear Kolomogorov scaling in the inertial range.
We argue that the underlying reason for having no intermittency, is the strong correlations between the velocity fields at different shells.
We analyze the tails of velocity distribution functions, which offer new insights to the structure of fluid turbulence.

\end{abstract}

\maketitle

\section{Introduction}

Shell models of turbulence are simplified mathematical models of fluid turbulence,
that reduce the complexity of Navier-Stokes equations by considering the dynamics of the energy at
shells of discrete wave numbers. The shells correspond to different scales of turbulence dynamics, 
and the shell models study the interaction between them.
The shell models include nonlinear interactions between the shells, which mimic the complex interactions that occur in actual turbulent flows, yet simplify the turbulence problem by only considering the exchange of energy between neighboring shells.
Shell models capture the energy cascade of turbulence as a transfer of energy between adjacent shells, and 
allow for a simplified study of the statistical properties of turbulence including intermittency and scaling at the inertial range.

In \cite{creswell2024anomalous} (See also \cite{de2024extreme} for the computation of the leading twenty five moments.) we performed a precision analysis of shell model of a complex velocity field in the steady state turbulent regime, including a calculation of the leading hundred anomalous scaling exponents, 
the probability distribution function of the magnitude and phase of the velocity, the correlations among them
at different shells and the tail of velocity distribution function. This precision analysis allowed us to establish
a linear scaling of the high moments that differs from Kolomogorov's, as well as gaining
new insights to the structure of the velocities probability distribution functions.
The aim of this work is to perform a precision analysis of the direct and inverse cascades  
in shell models of turbulence, where the velocity field is a real-valued function. 

The real-valued shell model \cite{DN} is a set of coupled non-linear equations: 
\begin{eqnarray}
    \frac{du_n}{dt} + \nu k_n^2 u_n = \alpha \qty(k_n u_{n-1}^2 - k_{n+1} u_n u_{n+1}) + \nonumber\\
    + \beta \qty(k_n u_{n-1} u_n - k_{n+1} u_{n+1}^2) + f_n \ ,
    \label{eqmodel}
\end{eqnarray}
where the numbers $u_n(t)$ are real-valued velocity fields, and $n = 1, 2, \dots$ is called
the shell index.
$\alpha$, $\beta$ are real-valued constants, $k_n$ are wavenumbers obeying $k_n = k_0 \lambda^n$, where $k_0 > 0$ is the reference wave number, and $\lambda$ is a constant which we will set to $2$.
$\nu$ is a viscosity parameter, and 
$f_n$ is an external forcing, that will be taken to be a Gaussian random noise.
The random forcing will be applied to the first shells (IR) for the direct cascade, and to the last shells (UV) for the inverse cascade. The energy $E= \frac{1}{2}\sum_n u_n^2$ is conserved
separately for each of the $\alpha$ and $\beta$ interaction terms in (\ref{eqmodel}).

As we will show, the shell model reaches a steady state, and exhibits scaling at the inertial range of scale:
\begin{equation}
    \label{eq:model}
    \expval{(u_n)^p} \propto k_n^{-\zeta_p} \ ,
\end{equation}
where the average is taken over the space of solutions in the steady state turbulent regime.
The scaling exponents $\zeta_p$ are generally expected to be universal and independent of the force and viscous structures.
We will determine accurately these scaling exponents up to $p=100$, 
the marginal probability distribution functions of the velocity field at different shells, as well as the correlations between
 the velocities at different shells.
An important feature of this shell model turbulence, which we will observe is the strong correlation between the velocity at different shells, which holds for both the direct and inverse cascades. Thus, in contrast to the anomalous scaling in the complex shell model \cite{creswell2024anomalous}, here we will find a Kolmogorov linear scaling 
\cite{Kolmogorov1941} in both the direct and the inverse cascades.

\subsection{Time-domain data analysis}

We will be using simulations in which equation \eqref{eq:model} is solved numerically. 
For the direct cascade, the forcing term $f_n$ will be taken to be zero in all shells except the second, where it is a random correlated Gaussian noise variable.
45 shells will be used and the viscosity parameter $\nu$ in \eqref{eq:model} will be chosen to be $1 \times 10^{-6}$, which is a good balance for stability and integration time (smaller values of $\nu$ require extremely small time steps and very slow integration). For the results presented below, we will take $\alpha = \beta = 1$ \footnote{See the discussion section for a brief summary of the results for other values of $\alpha$ and $\beta$.}. 

For the inverse cascade, the forcing term is transferred to the last shell, $n = 44$. 
At the same time, we will add a dissipation term in the second shell, proportional to $-u_2$ with a coefficient of around $0.01$, which is found to give good performance.
The dissipation term ensures that a steady-state is reached.
Alternatively, a boundary condition can used, having the same effect.

The total amount of data that we simulated is $7 \times 10^6 \times T$, where  $T = 1 / (k_1 \times \expval{|u_1|})$ is the turnover time. Note, that this is substantially larger than the time step sizes used for the actual integrating of the differential equation.
The velocities $u_n(t)$ have non-constant profiles, and have a non-trivial time evolution. They take both positive and negative values, but primarily positive. In fact, some shells hardly take negative values. 
It is also apparent that there are strong correlations between the shells velocities (See correlation matrix calculation presented below).

The paper is organized as follow.
In section II we will analyse the direct cascade, and in section III the inverse cascade. We will present the scaling exponents $\zeta_p$ (\ref{eq:model}) up to $p=100$ and establish accuratley the Kolomogorov linear scaling in both cascades (Fig. \ref{M}). In fact, to the best of our knowledge, this is the first model of turbulence that exhibits Kolmogorov
scaling in the direct cascade. This is in tension with Landau's argument about the absence of scale invariance 
in the range of scales smaller than the forcing scale \cite{Frisch}.

We will study the probability distribution function for the velocity moments and show that it realizes Kolomogorov's scaling hypothesis, namely it is a function of the variable $u_n k_n^{-\frac{1}{3}}$ (Fig. \ref{PDF}), where $n$ is the shell index. Moreover, similarly to the complex shell model \cite{creswell2024anomalous}, the probability distribution function exhibits an extremely fast decaying tail, and a localized contribution 
to the high moments (Fig. \ref{P50} and Fig. \ref{Tail}).
Thus, the scaling exponents of the high moments are determined by the maximal velocity of each shell
(Fig. \ref{MV1} and Fig. \ref{MV2}).
Unlike the complex model \cite{creswell2024anomalous}, we find that the velocities at the different shells in the real model
are strongly correlated (Fig. \ref{CD} and Fig. \ref{CI}), which provides an explanation for having no intermittency.
While, the direct and inverse cascade exhibit the same universal scaling, we find that they can be differentiated
by the amplitude of the energy power spectrum (Fig. \ref{Amp}).

Section IV will be devoted to a discussion.

\section{Direct Cascade}

\begin{figure}
    \centering
    \includegraphics[width=0.9\linewidth]{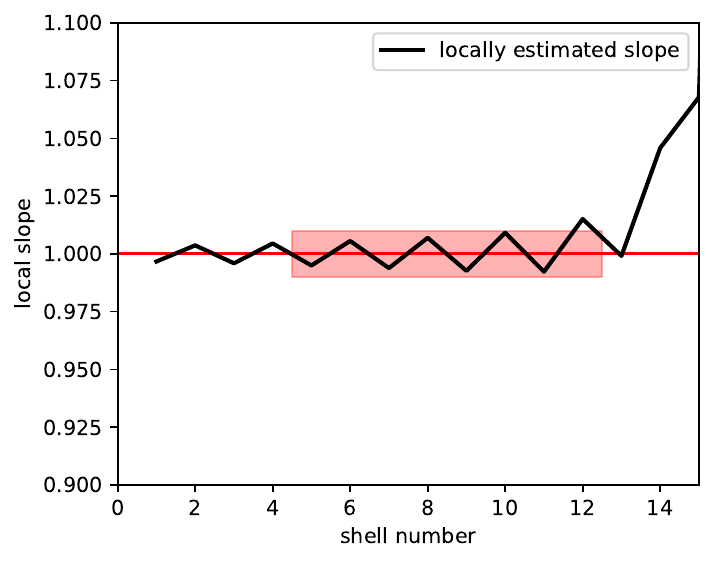}
    \caption{Determination of the inertial range. The 3rd moment can be estimated locally by a pair of adjacent shells. We consider the inertial range to end when this estimate departs from the theoretical value by more than 1\%. Note there is a small even--odd oscillation in the slopes. Furthermore, because the 3rd exponent here is calculated from only two shells, the errors are much higher than the actual exponent in Table 1, calculated by a fit over all shells.}
    \label{I}
\end{figure}

In this section we will analyze the direct cascade.
The forcing term $f_n$ will be taken to be zero in all shells except the second, where it is a random correlated Gaussian noise variable.
The inertial range is from shells 5 to 11, as determined by the third moment $\zeta_3$ in Fig. \ref{I}.

\subsection{Scaling exponents up to $p = 100$}

\begin{figure}[h!]
    \centering
    \includegraphics[width=0.5\textwidth]{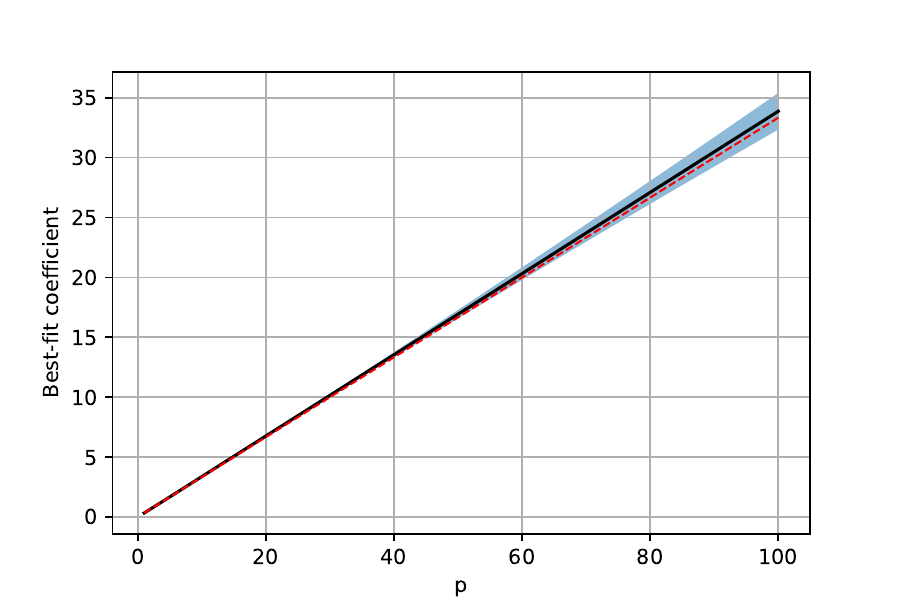}
    \caption{Scaling exponents $\zeta_p$ (\ref{eq:model}), up to $p=100$. Black is the real data, blue is the 99\% standard error region, and the red dashed line is $p/3$, which is within error bars.
    The numerical values of $\zeta_p$ are given in the table.}
    \label{M}
\end{figure}

The scaling exponents (\ref{eq:model}) exhibit the Kolmogorov's linear scaling, without intermittency.
Interestingly, this is the first model of direct cascade turbulence such a scaling.
In Fig. \ref{M} we plot the scaling exponents $\zeta_p$ (\ref{eq:model}), up to $p=100$.
Black is the real data, blue is the 99\% standard error region, and the red dashed line is $p/3$, which is within error bars. The numerical values of $\zeta_p$ are given in the table.

\begin{table}[ht!]
    \centering
    \begin{tabular}{|c|c|}
    \hline
    $p$ & $\zeta_p$  \\
    \hline
    \hline
1 & $0.333 \pm 0.00004$\\
2 & $0.667 \pm 0.00004$\\
3 & $1.000 \pm 0.00001$\\
4 & $1.333 \pm 0.0001$\\
5 & $1.667 \pm 0.0007$\\
6 & $2.005 \pm 0.002$\\
7 & $2.334 \pm 0.004$\\
8 & $2.663 \pm 0.008$\\
9 & $3.012 \pm 0.012$\\
10 & $3.341 \pm 0.018$\\
11 & $3.701 \pm 0.024$\\
12 & $4.040 \pm 0.031$\\
13 & $4.349 \pm 0.039$\\
14 & $4.728 \pm 0.047$\\
15 & $5.067 \pm 0.057$\\
16 & $5.406 \pm 0.068$\\
17 & $5.745 \pm 0.079$\\
18 & $6.084 \pm 0.092$\\
19 & $6.423 \pm 0.105$\\
20 & $6.762 \pm 0.119$\\
21 & $7.101 \pm 0.135$\\
22 & $7.440 \pm 0.151$\\
     \hline 
    \end{tabular}
    \begin{tabular}{|c|c|}

    \hline
    $p$ & $\zeta_p$  \\
    \hline
    \hline
   23 & $7.779 \pm 0.168$\\
24 & $8.118 \pm 0.186$\\
25 & $8.458 \pm 0.204$\\
26 & $8.797 \pm 0.224$\\
27 & $9.135 \pm 0.244$\\
28 & $9.474 \pm 0.266$\\
29 & $9.813 \pm 0.288$\\
30 & $10.152 \pm 0.311$\\
35 & $11.847 \pm 0.440$\\
40 & $13.541 \pm 0.589$\\
45 & $15.236 \pm 0.761$\\
50 & $16.930 \pm 0.953$\\
55 & $18.624 \pm 1.167$\\
60 & $20.318 \pm 1.402$\\
65 & $22.012 \pm 1.659$\\
70 & $23.706 \pm 1.938$\\
75 & $25.400 \pm 2.238$\\
80 & $27.094 \pm 2.560$\\
85 & $28.788 \pm 2.903$\\
90 & $30.482 \pm 3.268$\\
95 & $32.176 \pm 3.655$\\
100 & $33.870 \pm 4.063$\\ \hline 
    \end{tabular}
\end{table}

\subsection{Marginal distributions}

\begin{figure}[h!]
    \centering
        \includegraphics[width=0.8\linewidth]{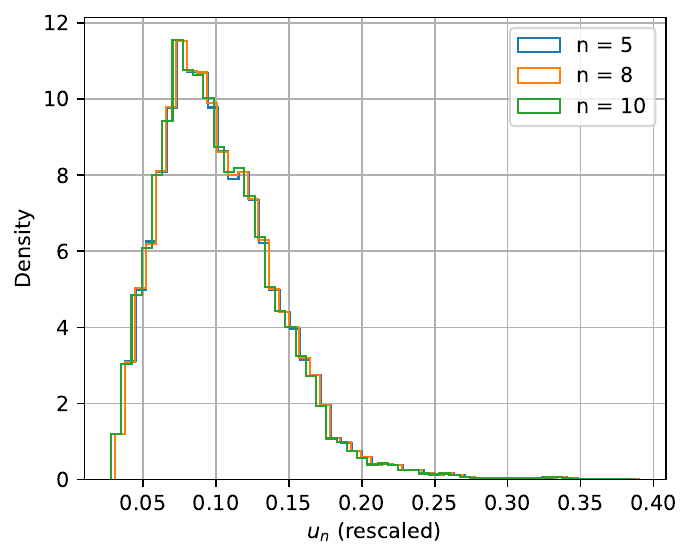}
            \caption{Marginal velocity probability distribution functions, after rescaling  $u_n$  by a factor of $k_n^{1/3}=2^{n/3}$. Within the inertial range, the marginals are related by this scaling, which breaks down outside the inertial range. 
            This realizes Kolomogorov's scaling hypothesis, where the velocity
            probability distribution is a function of the variable $u_n k_n^{-\frac{1}{3}}$.
        }
    \label{PDF}
\end{figure}

As we showed above, the moments of velocity marginal probability distribution functions of the shell velocities $u_n$ 
(\ref{eq:model}) exhibit Kolmogorov's linear scaling. The probability distribution functions themselves realize 
Kolomogorov's scaling hypothesis and are functions of the variable $u_n k_n^{-\frac{1}{3}}$ in the inertial range, as shown in Fig. \ref{PDF}. 

\begin{figure}
    \centering
    \includegraphics[width=0.5\linewidth]{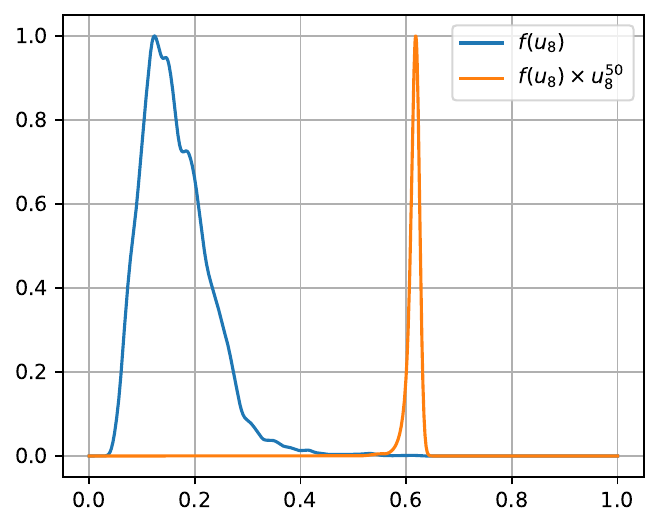}
    \caption{Integrand of the $p = 50$ moment, which is localized deep in the tail of the distribution function.
    We see a localized contribution 
to the high moments.
    Note, that the curves in this plot are normalized to have the same maximum for visibility.}
    \label{P50}
\end{figure}

\begin{figure}
    \centering
    \includegraphics[width=0.5\linewidth]{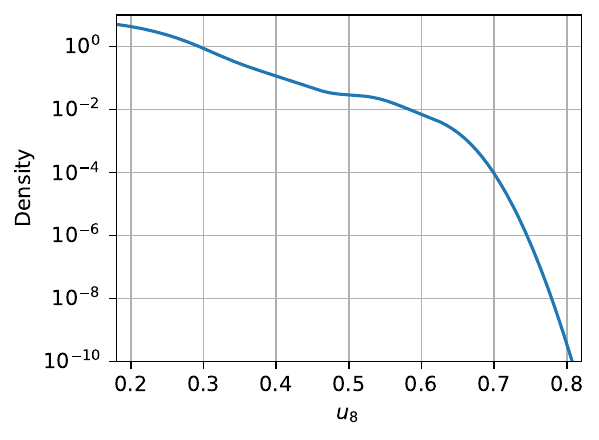}
    \caption{Tail of the $n = 8$ shell distribution function, zoomed in and with logarithmic axis.
     The probability distribution function exhibits an extremely fast decaying tail.
 }
    \label{Tail}
\end{figure}

The tail of the probability distribution function shows an extremely fast decay, as in Fig. \ref{Tail}, which is the reason why we are able to calculate accurately its moments up to such high orders. 
Indeed, the integrands of the high moments are localized deep in the tail of the distribution function as in Fig. \ref{P50},
and their dominant contribution comes from the maximal velocity.

\begin{figure}[h!]
    \centering
    \includegraphics[width=0.8\linewidth]{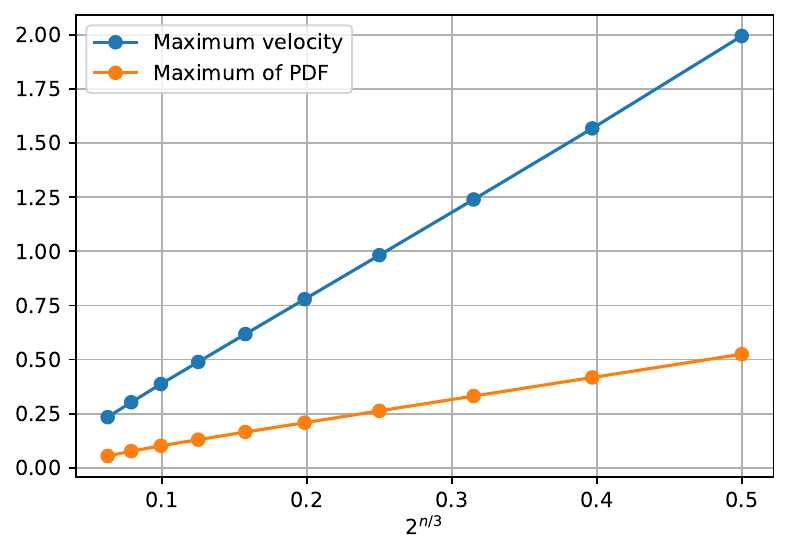}
    \caption{Maximal velocity in each shell (blue), and peak velocity of the PDF (orange), in the direct cascade. The x-axis has been rescaled in terms of $2^{n/3}$, resulting in linear dependence of the maxima on this variable.
    The high moments are dominated by the maximal velocity, hence exhibiting $p/3$ scaling exponents. Such as structure can explain a linear scaling of high moments even in the presence of intermittency \cite{creswell2024anomalous}.
    }
    \label{MV1}
\end{figure}

\begin{figure}[h!]
    \centering
    \includegraphics[width=0.8\linewidth]{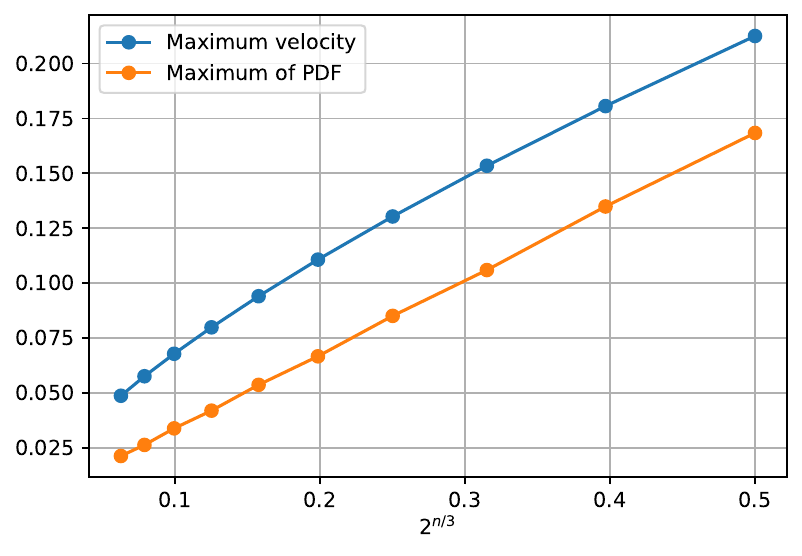}
    \caption{Maximal velocity in each shell (blue), and peak velocity of the PDF (orange), for the inverse cascade. The x-axis has been rescaled in terms of $2^{n/3}$, compare to figure~\ref{MV1}.
    }
    \label{MV2}
\end{figure}

The maximal velocity in each shell, and peak velocities of the probability distribution functions
are plotted in Fig. \ref{MV1}, where the horizontal axis has been rescaled by $2^{n/3}$, which is consistent with having Kolomogorov's scaling and no intermittency.
Note, that such as structure explains a linear scaling of high moments even in the presence of intermittency, 
as in the complex velocity shell model \cite{creswell2024anomalous}.

\subsection{Correlation matrix}

\begin{figure}[h!]
    \centering
    \includegraphics[width=0.8\linewidth]{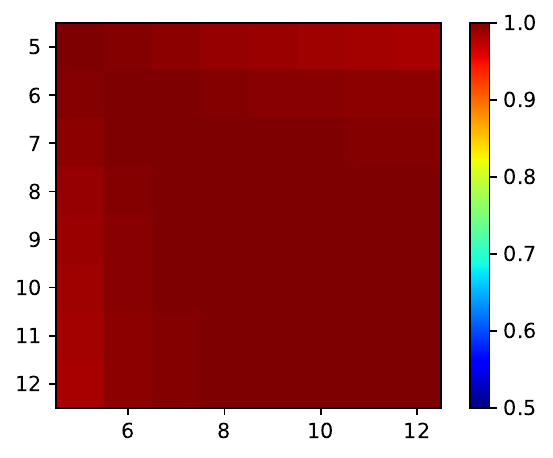}
    \caption{Direct Cascade: the correlation structure between the velocities at different shells $\langle u_nu_m\rangle$,
    within the inertial range. We see that the shells are very strongly correlated, generally above 90\%.}
    \label{CD}
\end{figure}

The joint probability distribution function of the shell velocities $u_n$, encodes additional information about the structure of turbulence.
Of particular importance is second moment of the distribution that encodes the correlation between the velocites
at different shells $\langle u_nu_m\rangle$. In Fig. \ref{CD} we plot this correlation  within the inertial range, and as be seen 
the shells are very strongly correlated, generally above 90\%. This strong correlation provides an explanation for the absence of intermittency in the shell model of real velocity field, and can be contrasted with the 
the intermittent complex velocity shell model, where there are weaker correlations  \cite{creswell2024anomalous}.

\section{Inverse cascade}

In order to create the inverse cascade we change the energy injection shells from $n=1$ to $n = 44$, and
to obtain a steady state we add a dissipation term at small $n$. 
As the direct cascade, also the inverse cascade inertial range turbulence exhibits Komologorov's scaling, and in the following we will outline the similarities and the differences between the two cascades. In general, the two cascades have a very similar universal structure in the inertial range, but they differ significantly in the non-universal behaviour outside this range.

\begin{figure}[h!]
    \centering
    \includegraphics[width=0.8\linewidth]{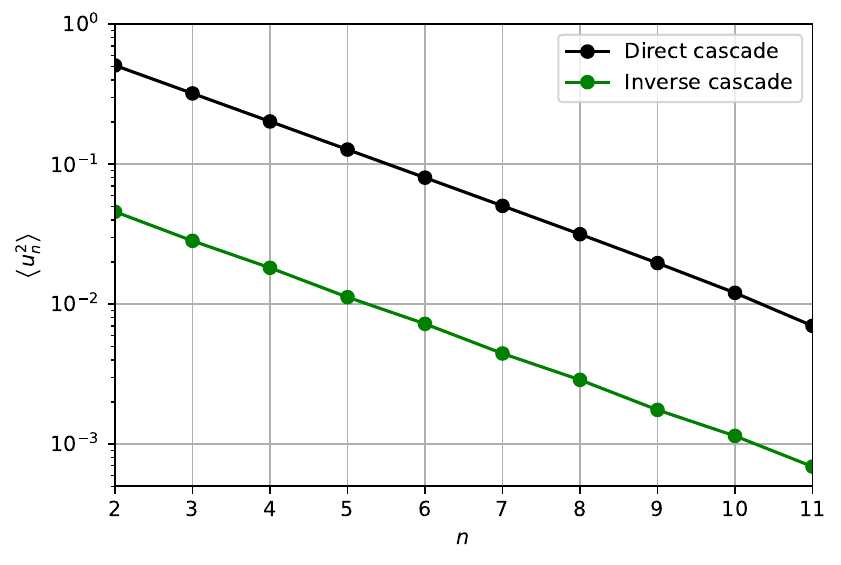}
    \caption{Energy spectrum for the inverse cascade (green) compared to the direct cascade for $p=2$. The main features are (which are common for other $p$, odd and even): the inertial range is slightly extended, but the slope, i.e. the scaling exponent is exactly the same, in this case we have $k_n^{-2/3}$. The overall coefficient is shifted. The inverse cascade spectrum has enhancement at high $n$, while the direct cascade spectrum falls monotonically for increasing $n$.}
    \label{Amp}
\end{figure}
\begin{figure}[h!]
    \centering
    \includegraphics[width=0.8\linewidth]{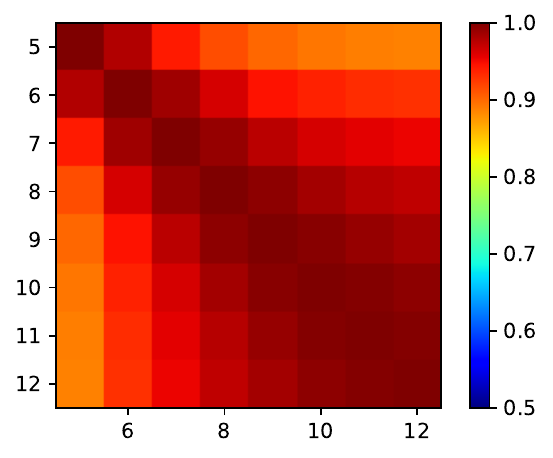}
    \caption{Inverse Cascade: the correlation structure between the velocities at different shells $\langle u_nu_m\rangle$,
    within the inertial range. We see that the shells are very strongly correlated.}
    \label{CI}
\end{figure}

The inverse cascade inertial range is slightly extended compared to the direct cascade,
while the scaling exponents $\zeta_p$ of the velocity moments (\ref{eq:model}) do not change and are described accurately by 
Fig. \ref{M} and by the associated table.
The overall coefficient of the moment, which is non-universal, is reduced compared to the direct cascade by  
approximately the factor $\frac{1}{10}$. The velocities marginal probability distribution functions have the same shape as in the direct cascade, but are reduced in scale by the same scale factor.
In Fig. \ref{Amp} we compare the energy spectrum ($\zeta_2$) of the inverse cascade with that of the direct cascade.
We see the same $k_n^{-2/3}$ scaling exponent, while the overall coefficient is shifted. 
In Fig. \ref{CI} we show the correlation structure between the velocities at different shells $\langle u_nu_m\rangle$, within the inertial range, and we see that as in the direct cascade, the shells are very strongly correlated. 
    
\section{Discussion}

Our precision analysis of the direct and inverse cascades  
in shell models of turbulence, where the velocity field is a real-valued function, revealed a Kolmogorov linear scaling 
of the velocity moments with no intermittency.
The velocities distribution functions exhibit a scale invariant structure, and strong correlations between the velocity fields at different shells. The tails of velocity distribution functions decay extremely fast, thus allowing us to accurately take into account the rare events and calculate with precision the high moments. These offer new insights to the statistical structure of fluid turbulence, which may shed light on the universal statistical structure of incompressible fluid turbulence.

Since energy is independently conserved for the $\alpha$ and $\beta$ terms in (\ref{eqmodel}),
one can consider different choices of them.
The case $\alpha = \beta = 1$ yields both direct and inverse cascades, as described and characterized in the previous
sections.
We have also checked other combinations of the coefficients, and here summarize the qualitative behaviour.
When $\alpha = 1$ and $\beta = 0$, the direct cascade simplifies and localizes
on the time independent solution $u_n(t) = C n^{-1/3}$, for some constant $C$. Injection at high $n$ in this case is highly numerically unstable and an inverse cascade is not generated.
When $\alpha = 0$ and $\beta = 1$, no inertial range can be established for injection at low $n$. Meanwhile in this case the inverse cascade is present. Therefore, one may associate the direct cascade with the first term in equation~(\ref{eqmodel}), and the inverse cascade with the second term. Note, that all the cascades 
exhibit Kolmogorov linear scaling with no intermittency.

\section*{Acknowledgements}

This work is supported in part by the Israeli Science Foundation Excellence Center, the US-Israel Binational Science Foundation, the Israel Ministry of Science and the LMU-TAU International Research Grant.

\bibliography{shell.bib}

\end{document}